# Early Solar System irradiation quantified by linked vanadium and beryllium isotope variations in meteorites

Paolo A. Sossi[1]*, Frédéric Moynier[1,2], Marc Chaussidon[1], Johan Villeneuve[3], Chizu Kato[1], Matthieu Gounelle[2,4]

[1] Institut de Physique du Globe de Paris, Sorbonne Paris Cité, Université Paris Diderot, UMR CNRS 7154, F-75005 Paris, France

[2] Institut Universitaire de France, Paris, France

[3] Centre de Recherches Pétrographiques et Géochimiques–CNRS, BP 20, 54501 Vandoeuvre-lès-Nancy, France

[4] Institut de minéralogie, de physique des matériaux et de cosmochimie - UMR7590 Muséum National d'Histoire Naturelle, Paris, France

**X-ray emission in young stellar objects (YSOs) is orders of magnitude more intense than in main sequence stars[1,2], suggestive of cosmic ray irradiation of surrounding accretion disks. Protoplanetary disk irradiation has been detected around YSOs by *HERSCHEL*[3]. In our solar system, short-lived $^{10}$Be (half-life = 1.39 My[4]), which cannot be produced by stellar nucleosynthesis, was discovered in the oldest solar system solids, the calcium-aluminium-rich inclusions (CAIs)[5]. The high $^{10}$Be abundance, as well as detection of other irradiation tracers[6,7], suggest $^{10}$Be likely originates from cosmic ray irradiation caused by solar flares[8]. Nevertheless, the nature of these flares (gradual or impulsive), the target (gas or dust), and the duration and location of irradiation remain unknown. Here we use the vanadium isotopic composition, together with initial $^{10}$Be abundance to quantify irradiation conditions in the early Solar System[9]. For the initial $^{10}$Be abundances recorded in CAIs, $^{50}$V excesses of a few per mil relative to chondrites have been predicted[10,11]. We report $^{50}$V excesses in CAIs up to 4.4‰ that co-vary with $^{10}$Be abundance. Their co-variation dictates that excess $^{50}$V and $^{10}$Be were synthesised through irradiation of refractory dust. Modelling of the production rate of $^{50}$V and $^{10}$Be demonstrates that the dust was exposed to solar cosmic rays produced by gradual flares for less than 300 years at ≈0.1 AU from the protoSun.**

We have measured, for the first time, coupled $^{51}$V/$^{50}$V and $^{10}$Be/$^9$Be ratios in six CAIs hand-picked from two CV3 chondrites (Allende and NWA 8616) and split into several fragments. Vanadium isotopic measurements were performed on a Thermo-Finnigan Neptune Plus Multi-Collector Inductively-Coupled-Plasma Mass-Spectrometer (IPGP, Paris, France). Vanadium isotope compositions are reported in delta notation with respect to the chondritic average, which is 1.50±0.35‰ lighter than the international Alfa Aesar standard (ref. 12):

$$\delta^{51}V = \left(\frac{(^{51}V/^{50}V)}{(^{51}V/^{50}V)_{chondrite}} - 1\right) \times 1000, \tag{1}$$

Five CAIs fragments were analysed for their $^{10}Be/^9Be$ composition using a CAMECA 1280 ion probe (CRPG, Nancy, France), following procedures described previously[13]. Rare-Earth Element (REE) concentrations were measured with an Agilent 7900 ICP-MS quadrupole (IPGP, Paris, France). All uncertainties quoted in the text and shown in the figures are 2SD.

Examination of the polished fragments of the CAIs shows that they are either fine-grained or coarse-grained (Supplementary Figures 1-5; Supplementary Table 1). The latter have flat chondrite-normalised rare-earth element (REE) abundances (Group V; Supplementary Figure 7; Supplementary Table 5) indicative of complete condensation of the REE budget of the solar nebula[14]. Contrastingly, the fine-grained CAIs have a Group II REE pattern (Supplementary Figure 7; Supplementary Table 5), which reflect equilibrium condensation from a gas previously depleted in the most refractory REE (with half-condensation temperatures, $T_c \approx 1650$ K[15]). The six CAIs show large variations in V contents and $\delta^{51}V$ values, which are correlated with the petrographic types (Fig. 1). Vanadium contents are high (151 – 516 ppm) and $\delta^{51}V$ values close to bulk chondrites in coarse-grained CAIs, while the fine-grained CAIs have chondritic vanadium contents (83 – 93 ppm) and show strongly negative $\delta^{51}V$ values, reaching -4.37±0.71‰ (Supplementary Table 2). All the CAIs studied show evidence for the presence of short-lived $^{10}Be$ when they formed (Fig. 2). The two coarse-grained CAIs have initial $^{10}Be/^9Be$ ratios of $(1.2\pm0.3)\times10^{-3}$ and Be contents of 0.015 and 0.026 ppm, whereas fine-grained CAIs have higher $^{10}Be/^9Be$, $(7.1\pm2.4)\times10^{-3}$, and less Be, 0.006-0.012 ppm (Supplementary Tables 3 and 4). The $^{10}Be/^9Be$ ratios in fine-grained CAIs are on the high end of those previously reported, which range from $3.6\times10^{-4}$ to $1.0\times10^{-2}$ (refs. 16, 17).

The observed variability of $^{10}Be$ abundance in the studied CAIs confirms $^{10}Be$ heterogeneity in the protoplanetary disk[5,18,19]. Although $^{10}Be$ production is commonly ascribed to solar cosmic ray irradiation of the accretion gas or dust, a fraction of the $^{10}Be$ could have been inherited from the pre-solar molecular cloud via trapping of galactic cosmic rays[19]. However, this process would result in a uniform $^{10}Be/^9Be$ ratio[19], a feature that is not observed (Fig. 1). Furthermore, in order to reach $^{10}Be/^9Be$ ratios of $10^{-3}$ in the model of ref. 19, the galactic cosmic ray flux was arbitrarily doubled with respect to observed fluxes, implying that it is even more implausible to obtain $^{10}Be/^9Be$ ratios as high as $10^{-2}$ by this process.

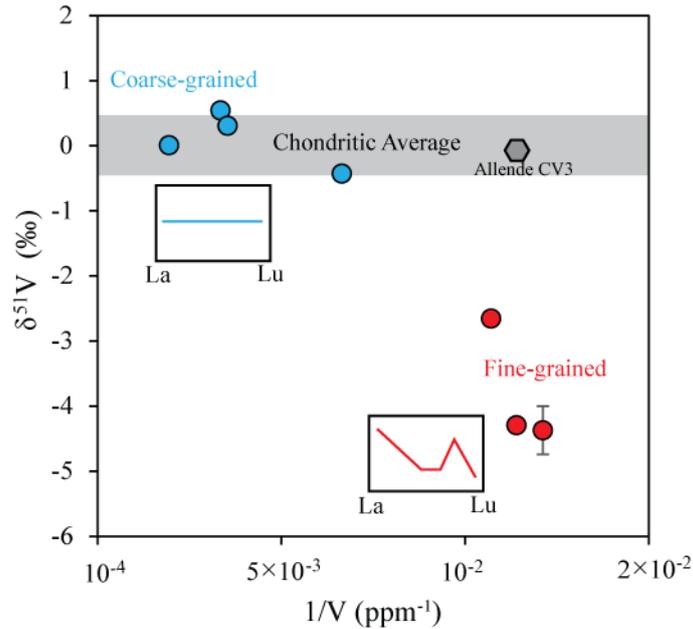

**Figure 1 Isotopic composition and concentration of V in CAIs** The variation in V isotope ratios of CAIs and the bulk Allende CV3 meteorite, expressed as $\delta^{51}V$ relative to chondritic meteorites (grey field) as a function of the inverse concentration of V (1/V). Fine-grained CAIs (red circles) have lower V abundances and $\delta^{51}V$ down to -4.5 ‰. Insets show the schematic REE patterns of the bulk CAIs.

As suggested by its co-variation with $^{10}Be$ (Fig. 1), the most likely source of the observed $^{50}V$ enrichment relative to the chondritic $^{51}V/^{50}V$ composition is solar cosmic ray irradiation[10,11]. However, other potential sources of vanadium isotope variation must be considered: (i) isotopic anomalies of nucleosynthetic origin and (ii) mass-dependent isotopic fractionation. Nucleosynthetic anomalies are unlikely to create a ~5‰ vanadium isotopic fractionation in CAIs since both vanadium isotopes are synthesised by a common nucleosynthetic process (explosive oxygen burning[20]), limiting isotopic heterogeneity endemic to the disk. Taking Ti isotopes as an analogy for vanadium, since $^{46}Ti$, $^{47}Ti$, and $^{49}Ti$ are synthesised in a similar manner[20], the small Ti isotopic anomalies in CAIs ($\leq 0.2‰$[21]) suggest any nucleosynthetic vanadium isotope anomalies should be insignificant at the per mil level. High-temperature, mass-dependent isotopic fractionation is documented in CAIs for some elements (Mg, Si, Ca, Ti, refs. 22–24), extending to both heavier and lighter compositions than bulk chondrites, and produced by both equilibrium and kinetic condensation/evaporation[25], in addition to electromagnetic sorting[26]. Any process, be it kinetic or at equilibrium, that causes light isotope enrichment relative to chondrites must also explain the combined systematics between $\delta^{51}V$, vanadium- and REE contents in the CAIs (Fig. 1). Where light isotope enrichment in CAIs has been observed for other isotope systems (*e.g.,* Ti isotope deviations of down to -1.5 ‰/amu relative to chondrites[24]), it is not simply correlative with CAI type (coarse- vs. fine-grained) or REE patterns (Group V or II). Furthermore, the $1/T^2$ dependence of equilibrium isotope fractionation renders it ineffective at the temperature of condensation of CAIs (1500 K). By contrast, kinetically-controlled processes would result in light or heavy isotope enrichment in CAIs, depending on whether vanadium mass transfer occurs into or out of the CAIs by partial condensation or partial evaporation, respectively. However, the group II REE patterns of fine-grained CAIs (Fig. 1) provide evidence for their formation under equilibrium conditions, rendering kinetic vanadium isotopic fractionation untenable in this context. Therefore neither kinetic nor equilibrium processes can produce the vanadium isotopic variations recorded in CAIs.

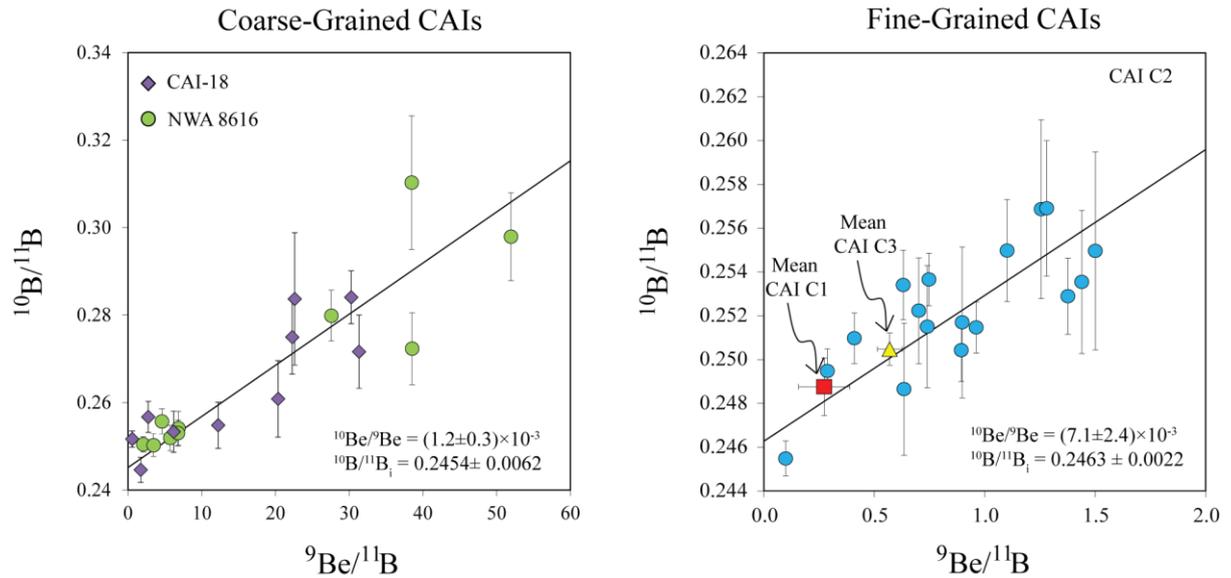

**Figure 2 Isochrons of Be-B data in fine- and coarse-grained CAIs** Boron-Beryllium isotope systematics of the two groups of CAIs, a) Coarse-grained and b) Fine-grained. Slopes of the isochrons of the coarse-grained CAIs yield $^{10}Be/^9Be \approx 10^{-3}$. Fine-grained CAIs are characterised by lower elemental Be/B, giving low $^9Be/^{11}B$ ratios and have an initial $^{10}Be/^9Be = 7.1(\pm2.4)\times10^{-3}$ significantly higher than the coarse-grained CAIs, $1.2(\pm0.3)\times10^{-3}$.

The following discussion is formulated on the basis that the major source of vanadium isotopic variation in the studied CAIs is due to irradiative production of $^{50}V$, as predicted by modelling[10,11]. The mass concentration, $X_{IRR}$ of any isotope produced by irradiation of a target T by a cosmic ray (CR) through any nuclear reaction, $i$, CR+T → $X_{IRR}$ can be written in the thin target approximation:

$$X_{IRR} = \sum_i \gamma_i X_{(T)i} \int \sigma_i(E) \frac{dF}{dE} dE, \qquad (2)$$

where dF/dE is the proton differential fluence, $\sigma_i(E)$ is the cross section of the considered nuclear reaction, $X_{(T)i}$ is the mass concentration of the target and $\gamma_i$ the relative abundance of the impinging cosmic ray relative to protons. The proton differential fluence can be written:

$$\left(\frac{dF}{dE}\right) = KE^{-p}, \qquad (3)$$

where $p$ is the spectral slope, which quantifies the relative abundance of high and low energy protons (with low p events corresponding to gradual flares[27] and high $p$ to impulsive flares), with

$$K = E_{10}^{(p-1)}(p-1)F_{10}, \qquad (4)$$

where $E_{10}$=10 MeV (ref. 11) and $F_{10}$ is the total fluence above 10 MeV experienced by the targets:

$$F_{10} = \int_{E_{10}}^{\infty} \frac{dF}{dE} dE. \qquad (5)$$

We consider cosmic rays composed of protons and helium nuclei and calculate $^{50}$V and $^{10}$Be production through a diversity of nuclear reactions (Supplementary Table 6). As nominal conditions, we use the ratios $^4$He/proton = 0.1 and $^3$He/$^4$He= 0.1 that characterise gradual flares from the present-day Sun[28]. Increasing the $^3$He/$^4$He to 1 – as would be the case for impulsive flares – does not change the production yields by more than 25%.

Models that favour irradiation of the nebular gas[8,29] predict homogeneous levels of $^{10}$Be in CAIs, which is contrary to our observations of high $^{10}$Be/$^9$Be (Fig. 1) and the low $^{10}$Be/$^9$Be ($\approx 4\times10^{-4}$) in the rare FUN (Fractionation and Unknown Nuclear effects) CAIs[18] and hibonite inclusions[30]. Because the initial $^{10}$Be/$^9$Be ratio varies sympathetically with the type of CAI (fine- vs. coarse-grained), the REE patterns and the $\delta^{51}$V, these observations support irradiation of the CAIs themselves rather than the gas from which they condensed. Therefore, we consider scenarios that involve the irradiation of solid targets (the CAIs or their precursors), leaving $p$ (the energy distribution spectral slope) and $F_{10}$ (the total fluence) as the two unknowns in the irradiation models.

The correlated $^{50}$V and $^{10}$Be excesses among coarse- and fine-grained CAIs can be reproduced by irradiation calculations (Fig. 3; Supplementary Figure 9) in two different irradiation scenarios. The first invokes differing degrees of irradiation of CAIs having their present chemical composition. Here, all CAIs have $\delta^{51}$V and $^{10}$Be/$^9$Be that define a near-constant spectral slope for cosmic rays ($p \approx 2.5$) with variable fluence, $F_{10}$ (Fig. 3; Supplementary Figure 9). Fine-grained CAIs experienced fluences of 2.5-6.5×10$^{18}$ cm$^{-2}$, whereas coarse-grained CAIs underwent lower fluence of $\approx 1\times10^{18}$ cm$^{-2}$. In this case, the higher product/target ratio (0.17 < V/Ti < 0.37 in coarse-grained CAIs compared with 0.03 and 0.15 for fine-grained CAIs), and, to a lesser extent Be/O in coarse-grained CAIs stifles the production of $^{10}$Be and particularly $^{50}$V (Fig. 3). An alternative scenario states that, since coarse-grained CAIs may form from melting of precursor grains similar to fine-grained CAIs[31], $^{50}$V and $^{10}$Be excesses in the precursor material could have been diluted by exchange with ambient nebular gas during re-melting. In order to produce the high vanadium (and, to an extent, Be; Supplementary Table 4) contents of the coarse-grained CAIs, the material with which they interact must necessarily be rich in V and Be, and hence refractory in nature. Mixing between fine-grained CAIs and an un-irradiated end-member can re-produce the observed trend in Fig. 3, but only when V/Be ratios in this refractory endmember are 40 times higher than chondritic, see Supplementary Figure 8. A refractory component with these characteristics is unlikely to exist in the primordial solar nebula, since both vanadium and beryllium are refractory lithophile elements with near identical half-condensation temperatures, ($T_c$ = 1429 K and 1452 K, respectively[15]) and thus difficult to fractionate from one another to the extent required. As such, this scenario is implausible, but importantly, the energy distribution spectral slope of the cosmic rays remains unchanged at $p = 2.5 \pm 0.4$ regardless of the scenario (resetting vs. chemical control). In the following discussion, we consider that $^{10}$Be and $^{50}$V excesses in CAIs were largely controlled by irradiation of refractory dust with different chemical composition by cosmic rays characterised by a variable fluence, $1 < F_{10} (\times 10^{18} \text{ cm}^{-2}) < 6.5$, and an energy distribution spectral slope of $p \approx 2.5$.

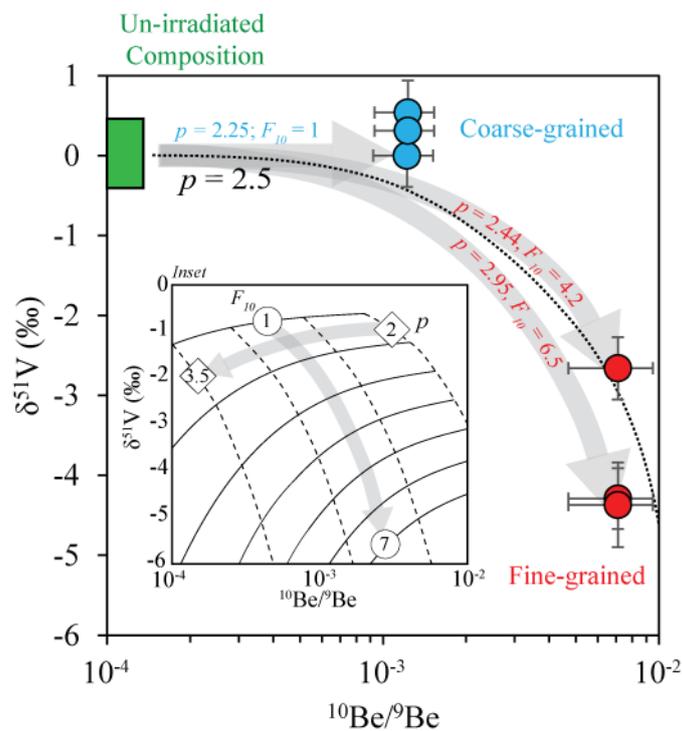

**Figure 3 Irradiation models re-producing the V- and Be isotope composition of CAIs** The evolution of $\delta^{51}$V and $^{10}$Be/$^9$Be in CAI precursor material during irradiation. Arrows in grey show the trajectory for individual CAIs as a function of cosmic ray flux ($F_{10}$; shown in reduced units; $F_{10}/10^{18}$ cm$^{-2}$) and energy distribution spectral slope ($p$), eqs. 2-5. *Inset:* An example grid for a chondritic precursor irradiated with a range of $p$ and $F_{10}$. The form of the grid changes with chemical composition (Supplementary Information).

Cosmic rays with characteristics derived from the calculations imply the predominance of gradual flares (low $p$) during CAI formation and exclude impulsive flares. Using these conditions and the appropriate cross sections (Supplementary Table 6), the coproduction of other short-lived radionuclides ($^{26}$Al, $^{41}$Ca, $^{53}$Mn) is calculated (Supplementary Table 7). We confirm that $^{26}$Al cannot be produced by irradiation to any meaningful extent[32], with calculated $^{26}Al/^{27}Al$ ratios an order of magnitude lower than the canonical ratio of $(5.23\pm0.13)\times10^{-5}$ (ref. 33). For irradiation of CAIs with a fluence, $F_{10}$ of $10^{23}$ cm$^{-2}$, abundances of $^{41}Ca$ and $^{53}Mn$ higher than the observed values were calculated by ref. 8, and used as a basis to exclude irradiation of the CAIs themselves. These high abundances arose because, in order to match the observed $^{10}Be/^{9}Be$ in CAIs, $F_{10}$ was arbitrarily increased to compensate for the otherwise inefficient production of $^{10}Be$ due to the high Be/O assumed[8]. In contrast, in our model using the low Be/O measured (Supplementary Table 4) and $F_{10}$ = 1-6.5$\times10^{18}$ cm$^{-2}$, calculated $^{41}Ca/^{40}Ca$ and $^{53}Mn/^{55}Mn$ are of the same order of magnitude as the handful of values measured in CAIs[34,35]. However, these values should be interpreted with caution, as Mn is prone to disturbance during thermal metamorphism of CAIs, and isotopic constraints on the $^{41}Ca$ abundance in CAIs are sparse. Nevertheless, our calculations allow for solar system irradiation of CAIs to have contributed to the inventory of these two short-lived radioactive elements.

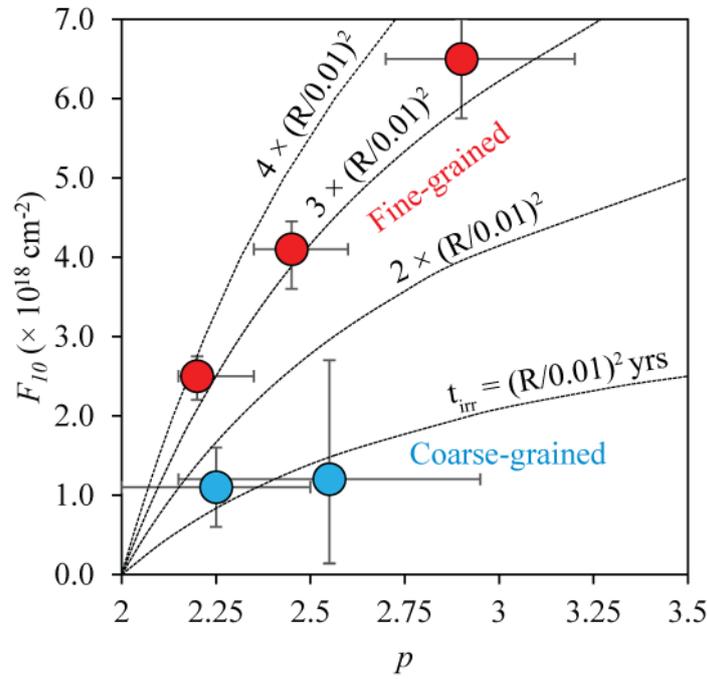

**Figure 4 Timescales of CAI irradiation** Variation of $p$ and $F_{10}$ calculated for CAIs. The irradiation time, in years, is proportional to $(R_{(AU)}/0.01)^2$. Taking R = 0.05 AU, all fine-grained CAIs (red circles) were irradiated for 75 years, corresponding to a residence time at the inner edge of the disk of ≈300 years, whereas the coarse-grained CAIs (blue circles) experienced shorter irradiation times.

With $p$ and $F_{10}$ in hand, the time for which the CAIs were irradiated can be estimated. The irradiation time ($t_{irr}$) is related to $p$ and $F_{10}$ by the expression:

$$t_{irr} = F_{10}\frac{(1-p)}{(2-p)}\frac{E_{10}}{L_{10}}(4\pi R^2), \tag{6}$$

where $L_{10}$ is the protosun proton luminosity[11], R the irradiation distance from the Sun and $E_{10}$ = 10 MeV. The observed average flare X-ray luminosity in YSOs is $L_X = 6\times10^{30}$ erg/s and $L_{10}=0.1 L_X$ (refs. 2,11). High-temperature mineralogy and geochemistry suggest that CAIs were formed at the inner edge of the disk, R < 0.1 AU[36]. Despite differing irradiation conditions, the fine-grained CAIs fall on a constant-time curve corresponding to $t_{irr} \approx 3\times(R/0.01)^2$ yr (Fig. 4). All fine-grained CAIs were therefore irradiated for a constant duration, a conclusion that is independent of the adopted heliocentric distance. For a range of reasonable heliocentric distances, the irradiation times vary from a few years (R = 0.01 AU) to 300 years (R = 0.1 AU) (Fig. 4). Coarse-grained CAIs record irradiation timescales slightly shorter than those for fine-grained CAIs, $t_{irr} \approx 1-2\times(R/0.01)^2$ yr, however, owing to their lack of $^{50}$V enrichment and more complex history that involved re-melting, this should be taken as a tentative estimate. Observations of YSOs indicate that they spend, on average, 1/6$^{th}$ of their time in flare activity[2], placing an upper limit of ≈2 kyr on the residence time of CAIs at the edge of the inner disk. This is slightly shorter than estimates for the time life of CAIs precursors that are on the order of 4 kyr, as derived from $^{26}$Al/$^{27}$Al systematics of bulk CAIs[37]. These timescales are consistent with the idea that CAIs were formed and irradiated in a hot zone proximal to the protosun for a brief period before being extracted to cooler, outer regions of the disk, thereby preserving their high temperature petrologic characteristics, and ceasing production of isotopes produced by irradiation. Irradiation constraints confine CAI formation and irradiation to periods where the Sun was a class 0-1 protostar as the class II (T-Tauri) phase lasts ≈ $10^6$ yr (e.g., ref. 38).

*Acknowledgements*

P.A.S. and F.M. are grateful to the European Research Council under the European Community's H2020 framework program/ERC grant agreement # 637503 (Pristine). M.C. and F.M. thank the LabEx UnivEarths (ANR-10-LABX-0023) and ANR Cradle (ANR-15-CE31-0004-01). M.G. and F.M. were supported by the Institut Universitaire de France. We appreciate for the assistance of Delphine Limmois and Laeticia Faure in the clean lab, and Julien Moureau on the Neptune at IPGP. We are grateful for the constructive and detailed comments of three anonymous reviewers, which resulted in a more complete framing of the debate and discussion of isotopic fractionation and existing irradiation models. We wish to acknowledge the editor, Luca Maltagliati, for his balanced and detailed editorial and scientific input.

*Author Contributions*

P.A.S. collected the V data, interpreted the results and wrote the paper. F.M. devised the project, interpreted the results and wrote the paper. M.C. devised the project, collected the Be/B data, interpreted the results and wrote the paper. J.V. collected the Be/B data. C.K. extracted and helped characterise the CAIs. M.G. devised the project, developed the modelling, interpreted the results and wrote the paper.

*Methods*

*Vanadium Isotope Measurements*

Vanadium has two stable isotopes, $^{51}$V (99.76 %) and $^{50}$V (0.24%). The high $^{51}$V/$^{50}$V ratio (≈400), combined with the presence of isobaric interferences, $^{50}$Cr (4.31 %) and $^{50}$Ti (5.34 %) on the minor V isotope pose problems for plasma mass spectrometric determination of V isotope ratios, necessitating a very thorough purification procedure. In order to reach sufficient levels of purity, $^{51}$V/$^{49}$Ti and $^{51}$V/$^{53}$Cr ≥ $10^5$ (ref. 1), a modified version of the seven-step column chromatography procedure of (*ref. 39*) is employed. Since V has only two isotopes, a double-spiking regime cannot be used and quantitative yields through the chemical purification are a pre-requisite in order to avoid the possibility of fractionation of V on the resin exchange sites. Finally, the small $^{50}$V isotope signal compared with that of $^{51}$V at a given concentration results in poor counting statistics using the typical $10^{11}$ Ω resistors. Rather than employing a $10^{12}$ Ω resistor on the more abundant $^{51}$V isotope, a $10^{12}$ Ω is used to enhance signal, and therefore counting statistics, on the minor isotope.

After hand-picking, CAIs were dissolved using a concentrated 1 mL HCl : 0.5 mL HF : 0.2 mL HNO$_3$ mixture at 140°C on a table-top hotplate. Following evaporation upon progressive addition of 15 M HNO$_3$ to decompose any insoluble fluorides, the samples were dissolved in 6M HCl in readiness for the column procedure. The first column, loaded with 1mL BioRad® AG1-X8 (200-400 mesh) removes Fe from the V fraction. Samples are loaded in 6M HCl which strips off V, whereas Fe remains bound on the resin to lower molarities. The second column, of identical column dimensions but loaded with TRU-Spec, removes Ti. Samples are loaded in 7M HNO$_3$, in which Ti remains on resin exchange sites, but V is eluted. The removal of Fe and Ti prevents the selective complexation of these metals by H$_2$O$_2$, which is employed in the third separation stage. Here, samples are dissolved and loaded in 0.01 M HCl + 3% H$_2$O$_2$ to oxidise V to its pentavalent form, forming the oxyperoxy (VO(O$_2$)$_2$)$^-$ species. The vanadium pentoxide species adheres to the resin (identical to column 1), while Cr and Ti are washed with 0.01 M HCl. Vanadium is converted to the cationic (VO(O)$_2$)$^+$ species through the addition of 1 M HCl, and is thus subsequently eluted. These three columns remove the majority of the matrix. Additional, 200 μL clean-up columns were employed to achieve the required V/Ti and V/Cr ratios. Columns 4 and 5 involve AG1-X8 (200-400) mesh in HF-form, ensuring Ti remains on the resin whereas V elutes in dilute HF-HCl (1:0.5 M) mixtures. Columns 6 and 7 are intended for Cr removal, and employ the same principle as in column 3. Following the four clean-up columns, samples were evaporated with 15 M HNO$_3$ and re-dissolved in 2% HNO$_3$.

Samples were run on the Thermo-Fisher Neptune Plus Multiple-Collector Inductively-Coupled-Plasma Mass-Spectrometer housed at the Institut de Physique du Globe de Paris, France. Solutions of 1 ppm V in 2% HNO$_3$ were introduced via a Scott Double Pass Cyclonic Spray Chamber to ensure high signal stability. Sensitivities in medium resolution (MR) mode using a 100 μL/min Teflon nebuliser, H skimmer cone and Jet sampler cone were 20 V/ppm for $^{51}$V. A $10^{12}$ Ω resistor on the $^{50}$V cup was used to improve the counting statistics on the minor isotope. Solutions were run with standard (Alfa Aesar) - sample bracketing correcting for mass bias and plasma drift.

All V isotope ratios are reported in delta notation:

$$\delta^{51}V = \left(\frac{(^{51}V/^{50}V)_{sample}}{(^{51}V/^{50}V)_{standard}} - 1\right) \times 1000. \quad \text{(M1)}$$

To ensure the isotope ratios were accurate and precise, the BDH solution standard, which has a composition of -1.19±0.12‰ relative to the Alfa Aesar standard[39,40] was run throughout the course of the analytical sessions, yielding an average -1.10±0.16‰ (2SD, *n* = 6; Supplementary Figure 6).

For samples, we make two important improvements to the measurement procedure that reduces the threshold ratios for $^{51}$V/$^{49}$Ti and $^{51}$V/$^{53}$Cr to be ≥ 5 × $10^4$. In the procedure of ref. 39, V isotope measurements are carried out in low resolution mode, meaning that $^{53}$Cr remains unresolved from polyatomic interferences, namely $^{40}$Ar$^{13}$C, and hence may result in spurious $^{53}$Cr/$^{50}$Cr ratios that are used to correct for the presence of $^{50}$Cr on $^{50}$V. We remedy this by running in medium resolution, whose resolving power; Δm/m ≈ 8000, is sufficient to distinguish between $^{53}$Cr and $^{40}$Ar$^{13}$C. Furthermore, the original study[39] employed an arbitrary instrumental mass fractionation factor, $\beta$, of -2 for Ti and Cr isotopes. Here, we directly measure $\beta$:

$$\beta = \frac{\ln(\frac{^{50}E_m}{^{i}E_m} / \frac{^{50}E_t}{^{M}E_t})}{\ln(M_{^{50}E} / M_{^{i}E})}. \quad (M2)$$

Where $i$ = 49 or 53, $M$ is the mass and $m$ and $t$ are the 'measured' and 'true' ratios, respectively. We find that $\beta_{Cr}$ = -1.57 in pure Cr solutions, and $\beta_{Ti}$ = -2.62 in pure Ti solutions. Both of these modifications result in more robust Ti and Cr corrections.

Data are listed in Supplementary Table 2. To assess the external reproducibility of the procedure, the CAI NWA8616 was dissolved twice (NWA8616-1 and -2), and passed through the entire chromatography procedure. The two dissolutions yielded values of -0.96±0.05‰ and -1.19±0.02‰, suggesting an external precision of ±0.15‰. Furthermore, the Allende CV3 chondrite, was also analysed, and gave a value of -1.63±0.10‰, in excellent agreement with the published value of -1.66±0.13‰[41]. Similarly, for the basaltic reference material, BHVO-2, the measured value, -1.05±0.07‰ is also within uncertainty of previous measurements (-0.89±0.08‰[41]). The additional uncertainty in the chondrite-normalised δ$^{51}$V comes from the propagation of the error of the average chondrite isotopic composition which we take to be at -1.50±0.35‰, relative to the Alfa-Aesar standard[42,43].

*Be-B Measurements*

Be-B concentrations and B isotopic ratios were measured at the national ion microprobe laboratory in CRPG-CNRS (Vandoeuvre-lès-Nancy, France) with a Cameca ims 1280 ion microprobe using the same procedure as described in detail in ref. 44. The samples were sputtered with a 12.5 kV O$^-$ primary beam and secondary $^9$Be$^+$, $^{10}$B$^+$, $^{11}$B$^+$ ions accelerated at 10 kV were analysed at a mass resolution M/ΔM = 2500 in monocollection mode using the central electron multiplier (EM) as a detector. High primary beam intensities of ≈ 30 nA were used producing craters of ≈ 30-40 µm diameter. Isotopic ratios were calculated from the ratios of average intensities measured at each spot after correction of these intensities for background (ranging from 0 top 0.11 count/sec, average 0.02 count/sec) and for dead-time of the electron multiplier (73 ns, the correction for dead-time being only significant for the GB4 standard). The internal GB4 silicate glass standard was used to determine the ion yield of Be relative to B (yield$_{Be/B}$ = ($^9$Be$^+$/$^{11}$B$^+$)$_{measured}$/($^9$Be/$^{11}$B)$_{true}$, with an average value of 3.17±0.13 at 2 sigma and a 2 sigma error of ±0.03 for 15 measurements) and the B instrumental isotopic fractionation ($\alpha_{inst}$= ($^{11}$B$^+$/$^{10}$B$^+$)$_{measured}$/($^{11}$B/$^{10}$B)$_{true}$, with an average value of 0.9771±0.0045 at 2 sigma and a 2 sigma error of ±0.0014 for 10 measurements). Errors on the final $^9$Be/$^{11}$B and $^{10}$B/$^{11}$B ratios were calculated by adding in quadratic way the internal errors of each spot (calculated from the counting statistic on $^9$Be, $^{10}$B and $^{11}$B) and the errors on yield$_{Be/B}$ and $\alpha_{inst}$. Data are presented in Supplementary Tables 3 and 4. Because the $^{10}$Be isochrons from the two coarse-grained CAIs (CAI18 and NWA 8616) were identical within errors, a single $^{10}$Be isochron was calculated for the two CAIs. For fine-grained CAIs, only C2 showed a small but significant range in $^9$Be/$^{11}$B allowing to determine a $^{10}$Be isochron (see Fig. 2). All the data for fine-grained CAIs C1 and C3 (10 spots each) were averaged to get bulk $^9$Be/$^{11}$B and $^{10}$B/$^{11}$B ratios for these two CAIs. These bulk compositions fall on the $^{10}$Be isochron calculated for CAI2. They were thus added to the data for CAI2 to determine a single $^{10}$Be isochron for the three fine-grained CAIs. The data were plotted using Isoplot v.4.15 (ref. 45) and employing 2SD uncertainties.

*REE Measurements*

A five % fraction of the initially dissolved CAI was conserved for trace element analysis. This aliquot was dried down in 15 M $HNO_3$ and taken up in 2 mL 2% $HNO_3$ in preparation for mass spectrometric analysis. Exactly 50 µL of this solution was diluted with 5 mL, resulting in dilution factors of ≈1000x. Detection limits are around 0.1 ppt, or approximately 0.1 ppb in the original sample. Standardisation was performed by correction to a series of synthetic solutions containing trace elements at the 500, 100, 10, 1, and 0.1 ppb levels. In addition, the USGS standard BHVO-2 was employed as an internal standard to account for oxide production on the REE masses. These solutions were introduced via a glass nebuliser to a Peltier-cooled glass spray chamber coupled to an Agilent 7900 quadrupole ICP-MS system at the Institut de Physique du Globe de Paris, France. Collision–reaction cell gases were not used in this study. Data are shown in Supplementary Table 5 and plotted, normalised to chondrite, in Supplementary Figure 7.

**Data Availability.** The data that support the plots within this paper and other findings of this study are available from the corresponding author upon reasonable request.

*Methods References*

**Supplementary Information** is linked to the online version of the paper at www.nature.com/nature.